\documentclass[usenatbib]{mn2e}
\usepackage[]{graphicx}

\title[CO wind in V4334 Sgr]{Warm high velocity CO in the wind of
Sakurai's Object (= V4334 Sgr)}

\author[S.~P.~S.~Eyres et al.]{S.~P.~S.~Eyres$^1$, T. R. Geballe$^2$,
V. H. Tyne$^3$, A. Evans$^3$, B. Smalley$^3$, \newauthor
H. L. Worters$^1$\\
$^1$ Centre for Astrophysics, University of Central Lancashire, Preston, PR1 2HE,
UK\\
$^2$ Gemini Observatory, Hilo, HI 96720, USA\\
$^3$ Department of Physics, School of Chemistry and Physics, Keele University, Keele, Staffordshire, ST5 5BG, UK\\
}

\date{Received; in original form}

\begin{document}
\label{firstpage}
\maketitle

\begin{abstract}
We present UKIRT UIST spectra of Sakurai's Object (=V4334~Sgr) showing
CO fundamental band absorption features around 4.7~$\mu$m. The
line-centres are at heliocentric radial velocity of
$-$170$\pm$30~km~s$^{-1}$. The number and relative strengths of the
lines indicate a CO gas temperature of $400\pm100$~K and CO column
density of 7$^{+3}_{-2}\times10^{17}$~cm$^{-2}$. The gas was moving
away from the central star at an average speed of
$\sim$290$\pm$30~km~s$^{-1}$ in 2003~September. The lines appeared
sometime between mid 1999 (well after the opaque dust shell formed)
and mid 2000 and may have been somewhat more blue--shifted initially
than they are now.
The observed CO velocity and temperature indicate the continued
presence of a fast wind in the object, previously seen in the He~{\sc
i} 1.083~$\mu$m line beginning just prior to massive dust formation,
and more recently in atomic and ionized lines. The dust continuum is
consistent with a temperature of 350$\pm$30~K, indicating
continued cooling of the shell. The similar CO temperature suggests
that the bulk of the CO absorption occurs just outside of the dust
continuum surface.
\end{abstract}


\section{Introduction}
\label{sec-intro}

Stellar evolution theory accounts reasonably well for the effects of
thermal pulses on the development of stars on the Asymptotic Giant
Branch.  Less well constrained are the effects of a late
thermal--pulse (LTP, once the star has shed its envelope) or
very--late thermal--pulse (VLTP), once the star has begun to descend
the white dwarf cooling track).  However, indications are that perhaps
10 to 20\%\ of low-- and intermediate--mass stars undergo LTP or VLTP.
It appears that the subsequent ``born--again'' evolution across the HR
diagram is short lived (perhaps a few centuries, but perhaps as short
as a few decades; see \citealt{Iben83,Iben96, Lawlor03}), and consequently very
few are observable at any one time.  Sakurai's Object (= V4334~Sgr) is
probably the first example of a VLTP observable with non--optical
instruments in the immediate post--flash epoch.

Sakurai's~Object was first identified as possibly undergoing a
helium--shell--flash on 1996~February~23 \citep{Nakano96, Benetti96},
but it is clear that it started to increase in brightness in mid--1994
\citep{Takemizawa97}.  In 1995 we began an infrared (IR) spectroscopic
monitoring programme using the United Kingdom Infrared Telescope
(UKIRT) and the cooled grating spectrometer CGS4 \citep{Mountain90};
more recently we have used the UKIRT Imaging Spectrometer UIST
\citep{Ramsay-Howat00}. Observations are carried out throughout the
star's observable period each year and the latest data, taken in 2003~September, are presented here.

\section{Observations}
\label{sec-observations}


Spectra of Sakurai's Object covering 1.4--5.3$\mu$m were obtained at
the United Kingdom 3.8 m Infrared Telescope on Mauna Kea on
2003~September~8 (UT), using the facility's imager/spectrometer UIST.
The instrument was configured with a 0.48\arcsec\ slit, providing
resolving powers of approximately 500 (600~km~s$^{-1}$) at
1.5--2.5$\mu$m, 1400 (214~km~s$^{-1}$) at 2.9--4.1$\mu$m, and 1200
(250~km~s$^{-1}$) at 4.5--5.2$\mu$m.  Spectra of early type stars were
obtained in order to correct for atmospheric transmission and provide
approximate flux-calibration. Where possible, hydrogen recombination
lines in the calibration stars were artificially removed prior to
ratioing.  This correction was not done at some wavelengths due to the
contamination of the hydrogen line profiles with strong telluric
features.  A Gaussian smoothing with FWHM of 1.5~pixels was applied to
the resultant spectra, degrading the resolving powers by $\sim$7\% from
those given above. Wavelength calibration in the M band was derived
from a quadratic fit to telluric absorption features and has a
$\pm3\sigma$ accuracy of $\pm$0.0004$\mu$m (25~km~s$^{-1}$).

\section{Results}
\label{sec-results}

The overall 1 to 5~$\mu$m spectrum in 2003~September remains dominated
by the dust continuum, as it had since 1999. Comparison with the 1-5um
spectrum obtained in 2002~July shows that the dust has continued to
cool (Tyne et al, in preparation).

\begin{figure*}
\includegraphics[angle=270,width=17.7cm]{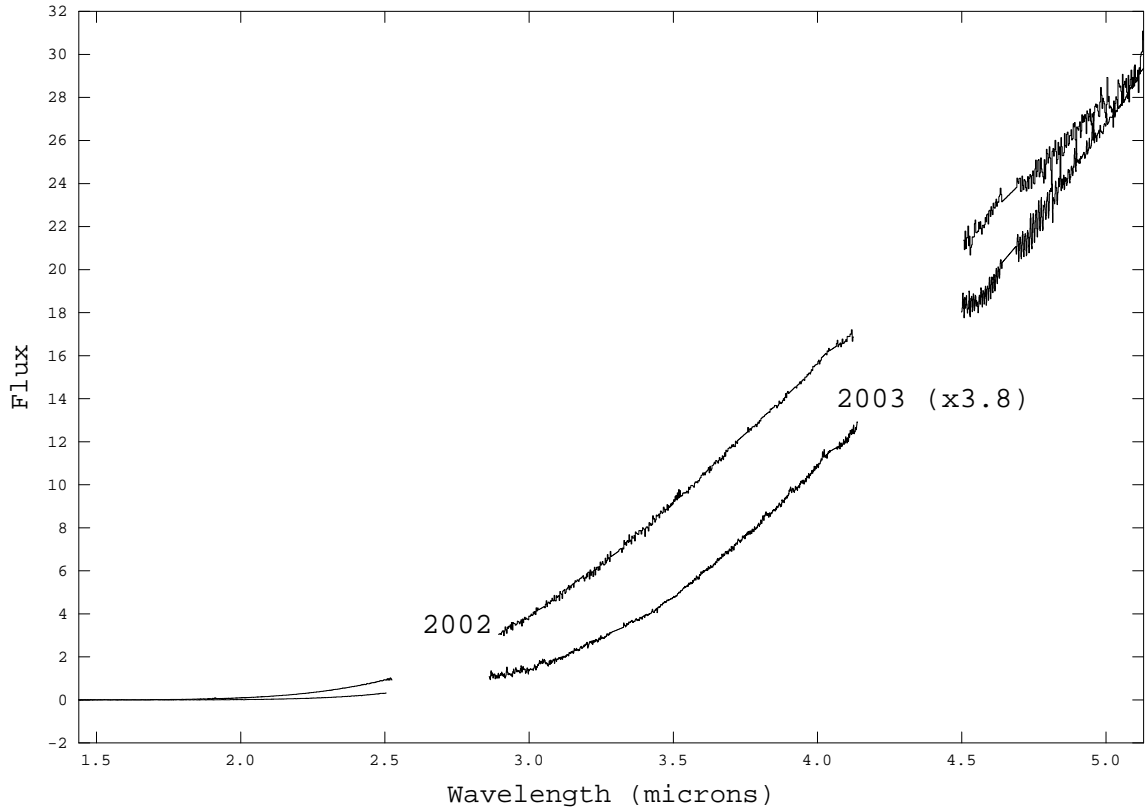}
\caption{UKIRT spectra from 2002~July~11 (upper line, Tyne et al. in
preparation) and 2003~September~8 (lower line). The latter has been
scaled by a factor 3.8 in flux to aid comparison of the continuum
form. Flux units are 10$^{-13}$~Wm$^{-2}~\mu$m$^{-1}$. The gaps are
due to gaps in the wavelength coverage.}
\label{fig-dust}
\end{figure*}

\begin{figure*}
\includegraphics[width=17.7cm]{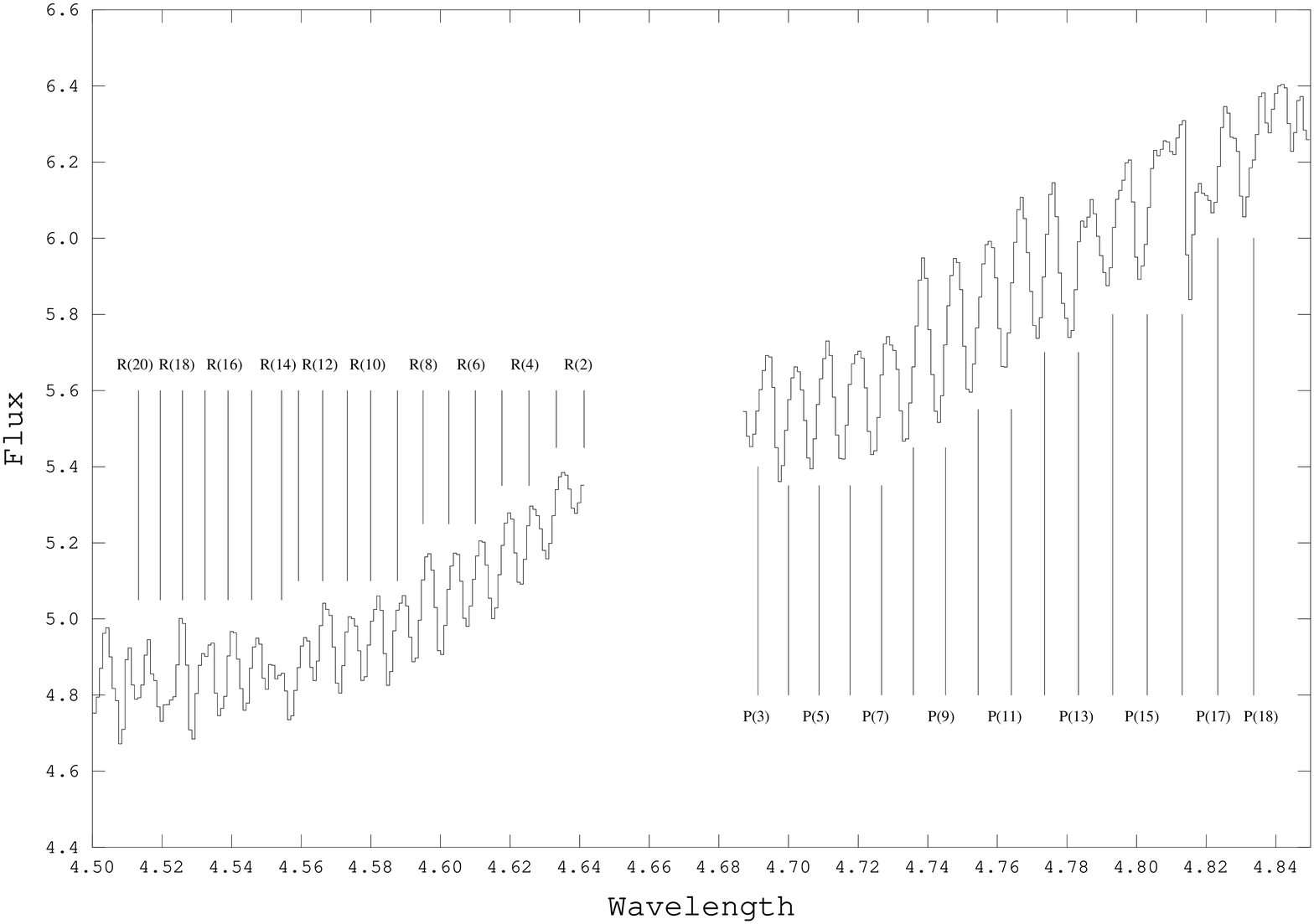}
\caption{UKIRT UIST spectrum from 2003~September~8 in the wavelength
range 4.45 to 4.95~$\mu$m . The CO fundamental band (1--0) lines are
clearly apparent in absorption. The rest wavelengths for the R and P
branch lines are marked by J--number. Flux units are
10$^{-13}$~Wm$^{-2}~\mu$m$^{-1}$. The gap at the centre of the
wavelength range is due to the removal of telluric features.}
\label{fig-spectra}
\end{figure*}

The central result of this paper is the detection of fundamental
vibration--rotation CO band lines around 4.7~$\mu$m in
2003~September. These are clearly visible in Fig.~\ref{fig-spectra},
and show a heliocentric average velocity--shift of
$-$170$\pm$30~km~s$^{-1}$. Essentially all the fundamental P--branch
lines from $J=3$ to $J=18$ are present ($J$ is the rotational quantum
number of the absorbing level). The spectral interval corresponding to
the lower end of the P branch has been removed due to the interference
of blended telluric and stellar absorption features. Similarly we see
the R--branch features from $J=2$ to 20 (with $J=0$ and 1 again lost
due to the calibration star features). The CO lines are unresolved in
the spectrum and therefore are narrower than about 200~km~s$^{-1}$
(FWHM).

Marginal evidence is present for detection of the $^{13}$CO
fundamental, whose band center is 4.77$\mu$m and whose line
wavelengths beat with those of $^{12}$CO. The strongest modulation in
the spectrum is near 4.75$\mu$m, corresponding to the wavelength of
strong $^{12}$CO and $^{13}$CO lines.  The small variation in the
modulation is consistent with a $^{12}$C/$^{13}$C ratio no less than 3
(in agreement with the recent determination of 4$\pm$1 by
\citet{Pavlenko04}) and we can rule out the lower half of the range of
values (1.5 -- 5) suggested by \citet{Asplund97, Asplund99}.  Since
the $^{12}$CO lines are not heavily saturated (otherwise the $^{13}$CO
lines would have a more pronouced effect on the spectrum), the FWHM's
of the CO lines must be no less than $\sim$25~km~s$^{-1}$.

We have re--examined spectra of Sakurai's Object dating back to
1999. We find that the CO lines are present but less prominent in the
2002 and 2000 spectra \cite[due largely to the lower resolution,
see][for a discussion of the earlier spectra]{Tyne02}, but noisiness
of the data makes this uncertain in the 2001 and most of the 1999
spectra. However a single M--band spectrum on 1999~May~4 (covering
only the region 4.588--4.748~$\mu$m) shows that the R(1) to R(7) and
the P(1) to P(9) lines are certainly absent.  Thus we can constrain
the appearance of these lines to after 1999~May~4, and state with
certainty that they were present by 2000~April~17. We infer that the
CO features appear to have been present for all but the early stages
of the current deep dust--induced optical dip.

The heliocentric radial velocity of Sakurai's~Object is
115~km~s$^{-1}$ \citep{Duerbeck96}.  Thus the observed CO is moving
away from the star at around 290$\pm$30~km~s$^{-1}$. The earlier data
from 2000~April~17 are consistent with a velocity relative to the star
of 360$\pm$60~km~s$^{-1}$, where the higher uncertainty is due to the
lower resolution and poorer signal--to--noise ratio. The difference
between the two values is not significant within the uncertainties.

Although we have shown that the $^{12}$CO lines are not heavily
saturated, their roughly comparable line strengths between $J=4$ and
$J=15$ are incompatible with optically thin lines existing within a
small range of temperatures. Because of this and the low spectral
resolution, the observed spectrum is not amenable to simple modeling
in order to obtain accurate values of temperature and column density.
However, crude estimates can be made as follows.  Beyond 4.81~$\mu$m
the spectrum is strongly affected by telluric absorption lines at many
wavelengths. The P(16) line is blocked from view. P(17) and P(18) are
clearly present, but the P(19) and higher lines are absent.  If the CO
temperature were higher than $\sim$500~K, the line intensities should
not drop off very rapidly beyond 4.83~$\mu$m and there is no reason
why we should not see more lines than we do.  A temperature less than
about 300~K would lead to very saturated low excitation lines. However
the low $J$ lines close to the band centre (4.67$\mu$m) are clearly
weaker than the higher $J$ lines. Thus we conclude that the CO
temperature is between 300 and 500~K, and adopt a value of
400$\pm$100~K. Our preliminary estimate the dust temperature, based on
the 1--5~$\umu$m spectrum (Fig.~\ref{fig-dust}) is 350$\pm$30~K. This
is consistent with the temperature of the CO and suggests that the CO
and the dust are at a similar distance from Sakurai's Object.

For the case of a P-branch CO fundamental band absorption line from the
ground vibrational state, the equation for the optical depth of a Gaussian
vibration-rotation line, given by Geballe et al. (1972) can be simplified
and transformed into the following expression for the column density of
CO:

\begin{equation}
N(\mbox{CO})~\approx ~2.9\times 10^{14}~\frac{\tau~T~\Delta v}{J}~e^{E/kT}~\mbox{cm}^{-2}
\end{equation}

\noindent where $T$ is the temperature of the CO, $\tau$ is the
optical depth, $\Delta v$ is the full width at half maximum in
km~s$^{-1}$, $J$ and $E$ the rotational quantum number and the energy
above ground of the absorbing level respectively. Using this we
estimate the column density of CO to be
$N~=~7^{+3}_{-2}\times10^{17}$~cm$^{-2}$, where the lower value
corresponds to the higher temperature. Here we have used the
(unresolved) P(15) line and assumed that it is optically thin so that
the product of the central optical depth and linewidth is constant at
a fixed value of $N$. If this assumption is incorrect or if there is a
significant column of cold CO (not well seen in these data), the
column density could be considerably larger.

\section{Discussion}
\label{sec-discussion}

\subsection{The origin of the CO}
\label{ssec-CO_origin}

There are several possible origins for the newly--discovered CO
features. We address each in turn.

\subsubsection{Interstellar CO}
\label{sssec-interstellar_CO}

The CO may be in the interstellar medium in the direction of
Sakurai's~Object. We can immediately rule this out. Interstellar CO
should have been visible with the same column all along (i.e. in the
1999~May spectrum). The number and relative strengths of the lines
(see \S~\ref{sec-results} above) indicates that the CO is too warm to
be interstellar. Finally the radial velocity suggests the CO is
associated with outflow from Sakurai's~Object.

\subsubsection{Recently--formed CO}
\label{sssec-new_CO}

The CO may have formed further from the star than the dusty material
shortly before detection (i.e. after 1999~May~4 but before 2000). However
the carbon--rich nature of the dust \citep{Tyne02} suggests there is
little oxygen in the region currently occupied by the CO. In addition
molecules must form before dust forms. It is more plausible that the
CO formed first, within the region now dominated by dust. Thus this
origin for the CO cannot be supported.


\subsubsection{CO formed in the dense wind}
\label{sssec-pre-existing_CO}

Neither of the previous suggestions for the origin of the dust is
tenable, and it appears most likely that the CO that is being observed
formed close to or in the outer layers of the central star prior to
2000.  In 1999~May the CO was not visible against the dust continuum,
but shortly thereafter it became detectable. Modeling of the evolving
dust continuum (Tyne et al, in preparation) suggests that the
continuum surface is expanding at about 50~km~s$^{-1}$. However, the
dust grains themselves may be moving at much higher speed than
this. The central object is probably continuing to generate a high
speed wind which, once it cools, is likely to form both molecules and
dust.  Clearly considerable gas is now outside of the continuum
surface and has been for some time (since 2000). An alternative model,
which seems more contrived, is that the dust shell and high speed gas
are separately produced entities and that the CO--containing gas
overtook the dust shell some time after the star was obscured.

%

\subsection{Fast wind}

We find CO absorption features consistent with a wind moving at
$\sim$300~km~s$^{-1}$. \citet{Eyres99} suggested a
$\sim$670~km~s$^{-1}$ wind explained a broad P~Cygni feature in helium
at 1.083~$\mu$m, with the breadth of the line consistent with material
moving at velocities down to $\sim$200~km~s$^{-1}$. We believe this
demonstrates that a fast wind continues to stream away from
Sakurai's~Object. The development of a fast wind following a VLTP and
the subsequent heating up of the cental star is a prediction of the
theory \citep{Iben83}. \citet{Kerber02} suggest that the central star
had entered this phase by 2001; we find that a fast wind has been in
evidence since 1998.

\section{Conclusion}
\label{sec-conclusion}

A recent spectrum of Sakurai's Object has clearly revealed the
presence of highly blue--shifted absorption lines of the fundamental
band of CO. The CO was first noted in a spectrum from
2003~September~8.  Re--examination of previous lower resolution
spectra shows that the CO features must have arisen between 1999~May~4
and 2000~April~17.  The blue--shift of the lines is consistent with
gas moving away from the central star at $\sim$300~km~s$^{-1}$. The
number and relative strengths of the lines suggest a CO gas
temperature of $400\pm100$~K and CO column density of
7$^{+3}_{-2}\times10^{17}$~cm$^{-2}$ (the lower value corresponding to
the higher temperature), and linewidths no greater than 25~km~s$^{-1}$
(FWHM). We rule out interstellar CO as the source of the absorption
lines, but the circumstellar CO could be formed shortly before
discovery or existed for some time prior to emerging from the dusty
shroud. 
We note that the CO velocity is additional evidence for the existence
of a fast wind in this object, although we cannot make a direct
connection with changes in the central star. Taking the 1--5~$\mu$m
spectrum overall we find that the dust continues to cool, and the
temperatures of the CO--bearing and dust--bearing materials are
similar. Hence the two components presumably lie at a similar distance
from the central star. As the CO is moving more rapidly than the dust
continuum surface, this suggests an ongoing wind replenishing the CO
in the region of the dust.

\section*{Acknowledgments}

\noindent HLW is supported by the University of Central
Lancashire. TRG is supported by the Gemini Observatory, which is
operated by the Association of Universities for Research in Astronomy,
Inc., on behalf of the international Gemini partnership of Argentina,
Australia, Brazil, Canada, Chile, the United Kingdom, and the United
States of America. The United Kingdom Infrared Telescope is operated
by the Joint Astronomy Centre on behalf of the UK Particle Physics and
Astronomy Research Council.

\bsp

\label{lastpage}

\begin{thebibliography}{}

\bibitem[\protect\citeauthoryear{Asplund et al.}{1997}]{Asplund97}
Asplund, M., Gustafsson, B., Lambert, D. L., Kameswara~Rao, N. 1997,
A\&A 321, L17

\bibitem[\protect\citeauthoryear{Asplund et al.}{1999}]{Asplund99}
Asplund, M., Lambert, D. L., Kipper, T., Pollacco, D., Shetrone,
M.D. 1999, A\&A 343, 507

\bibitem[\protect\citeauthoryear{Benetti, Duerbeck \&
Seitter}{1996}]{Benetti96} Benetti, S., Duerbeck, H. W., Seitter,
W. C. 1996, IAU Circular 6325



\bibitem[\protect\citeauthoryear{Duerbeck \& Benetti}{1996}]{Duerbeck96} Duerbeck, H. W., Benetti, S., 1996, ApJ, 307, L111






\bibitem[\protect\citeauthoryear{Eyres et al.}{1999}]{Eyres99} Eyres,
S. P. S., Smalley, B., Geballe, T. R., Evans, A., Asplund, M., Tyne,
V. H. 1999, MNRAS, 307, 11

\bibitem[\protect\citeauthoryear{Geballe, Wollman \& Rank}{1972}]{Geballe72}
Geballe, T. R., Wollman, E. R., Rank, D. M. 1972, ApJ, 177, L27





\bibitem[\protect\citeauthoryear{Iben, Kaler, Truran \& Renzini}{1983}]{Iben83}
Iben, I., Jr., Kaler, J. B., Truran, J. W., Renzini, A. 1983, ApJ,
264, 605

\bibitem[\protect\citeauthoryear{Iben, Tutukov \& Yungelson}{1996}]{Iben96}
Iben, I., Tutukov, A. V., Yungelson, L. R. 1996, ApJ, 456, 750



\bibitem[\protect\citeauthoryear{Kerber et al.}{2002}]{Kerber02}
Kerber, F., Pirzkal, N., De Marco, Orsola, Asplund, M., Clayton,
G. C., Rosa, M. R. 2002, ApJ, 581, 39


\bibitem[\protect\citeauthoryear{Lawlor \& MacDonald}{2003}]{Lawlor03}
Lawlor, T. M., MacDonald, J., 2003, ApJ, 583, 913

\bibitem[\protect\citeauthoryear{Mountain et al.}{1990}]{Mountain90}
Mountain, C. M., Robertson, D. J., Lee, T. J., Wade, R. 1990, in:
Instrumentation in astronomy VII; Proceedings of the Meeting, Tuscon, AZ

\bibitem[\protect\citeauthoryear{Nakano \& Sakurai}{1996}]{Nakano96}
Nakano, S., Sakurai, Y. 1996, IAU Circular 6322

\bibitem[\protect\citeauthoryear{Pavlenko et al.}{2004}]{Pavlenko04}
Pavlenko, Ya. V., Geballe, T. R., Evans, A., Smalley, B., Eyres,
S. P. S., Tyne, V. H., Yakovina, L. A., 2004, A\&A, in press

\bibitem[\protect\citeauthoryear{Ramsay-Howat et al.}{2000}]{Ramsay-Howat00}
Ramsay-Howat, S. K., Ellis, M. A., Gostick, D. C., Hastings, P. R., Strachan,
M. \& Wells, M. 2000, SPIE 4008, 1067


\bibitem[\protect\citeauthoryear{Takemizawa}{1997}]{Takemizawa97}
Takemizawa, K. 1997, VSOLJ Variable Star Bulletin, 25, 4


\bibitem[\protect\citeauthoryear{Tyne et al.}{2002}]{Tyne02} Tyne,
V. H., Evans, A., Eyres, S. P. S., Geballe, T. R., Smalley, B., Duerbeck H. W. 2002, MNRAS, 334, 875

\end{thebibliography}
\end{document}